\documentstyle[11pt,paspconf,epsf]{article}

\def\ltsima{$\; \buildrel < \over \sim \;$}
\def\simlt{\lower.5ex\hbox{\ltsima}}
\def\gtsima{$\; \buildrel > \over \sim \;$}
\def\simgt{\lower.5ex\hbox{\gtsima}}
\def\wave#1{$\lambda${#1}}

\def \oiii  {{\rm [O~III]}}
\def \lsun      {$L_{\sun}$}
\def \firr      {$f_{25}/f_{60}$}
\def \hb {H$\beta$}

\begin{document}

\title{Hidden Quasars in Ultraluminous Infrared Galaxies}

\author{H. D. Tran, M. S. Brotherton, S. A. Stanford, \& W. van Breugel}
\affil{Institute of Geophysics \& Planetary Physics,
    Lawrence Livermore National Laboratory, Livermore, CA 94550}

\begin{abstract}
Many ultraluminous infrared galaxies (ULIRGs) are powered by quasars hidden in 
the center, but many are also powered by starbursts. A simply diagnostic diagram is proposed that can identify obscured quasars in ULIRGs by their 
high-ionization emission lines (\oiii \wave 5007/\hb~\simgt~5), and ``warm'' IR
color ($f_{25}/f_{60}$~\simgt~0.25). 
\end{abstract}

Ultraluminous infrared galaxies (ULIRGs, $L_{IR} > 10^{12}$ \lsun) 
are an important constituent of our local universe, with luminosities 
and space densities similar to those of QSOs (Soifer et al. 1987). 
This led to the suggestion that ULIRGs could contain infant quasars 
enshrouded in a large amount of dust (Sanders et al. 1988). On
the other hand, they may also represent energetic, compact starbursts
(Condon et al. 1991). Understanding the dominant energy input mechanism in
these ULIRGs -- whether it is obscured quasars or intense bursts of
star formation --  has been the main issue concerning their nature.
In order to better understand the energy sources of ULIRGs and the
relationship between AGN and starburst activity, we started a 
spectropolarimetric survey to search for hidden broad emission lines
in a sample of ULIRGs that were identified in the cross correlations between 
the $IRAS~Faint~Source~Catalog$ ($FSC$) and those of the FIRST (Becker et al. 1995) and Texas (Douglas et al. 1996) radio surveys. 

Using the 10-m Keck II telescope, we obtained spectropolarimetric observations
of one ULIRG selected from a sample
identified in the FIRST-$FSC$ correlation (FF sources, Stanford et al. 1998), 
and two ULIRGs from the Texas-$FSC$ correlation (TF sources, Dey \& van Breugel 
1994). The infrared properties of these galaxies are listed in Table 1. 
\begin{table}
\caption{Infrared Properties}
\begin{center}\scriptsize
\begin{tabular}{lcccccccc}
Object & $z$ & $m$ & log($L_{IR}$/\lsun) & $f_{12}$ & $f_{25}$ & $f_{60}$ & $f_{100}$ & $f_{25}/f_{60}$ \\
\tableline
FF~J1614+3234 & 0.710 & 19.1 & 12.6 -- 13.2 & $<$ 0.065 & $<$ 0.055 & 0.174 & $<$ 0.54 &
 $<$ 0.31 \\
TF~J1020+6436 & 0.153 & 19.0 & 11.9 -- 12.1 & $<$ 0.095 & $<$ 0.062 & 0.86 & 1.24 & $<$ 
0.072 \\
TF~J1736+1122 & 0.162 & 18.0 & 11.8 -- 12.3 & $<$ 0.081 & 0.196    & 0.484 & $<$ 3.31 & 
0.404 \\
\end{tabular}
\end{center}
\end{table}
Our results show that only the high-ionization Seyfert 2 galaxy 
TF~J1736+1122 is highly polarized, displaying a broad-line spectrum visible 
in polarized light (Fig. 1a). The other two objects,
TF~J1020+6436 and FF~J1614+3234, exhibit spectra dominated by a population of 
young (A-type) stars similar to those of ``E + A'' galaxies. 
They are unpolarized, showing no sign of hidden broad-line regions.
The presence of young starburst components in all three galaxies indicates 
that the ULIRG phenomenon encompasses both AGN and starburst 
activity, but the most energetic ULIRGs do not necessarily harbor 
``buried quasars''. 
\begin{figure} 
\plottwo{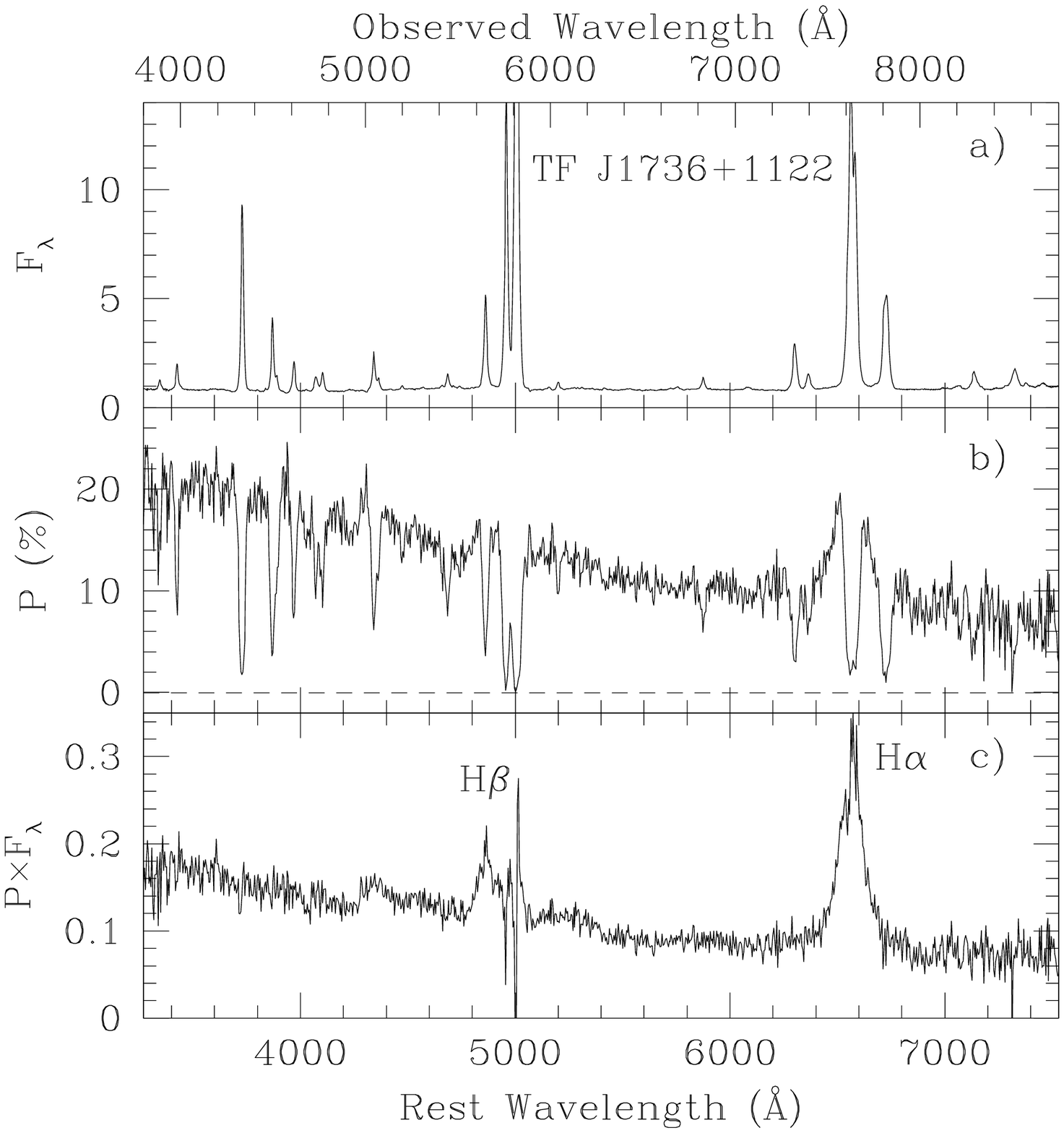}{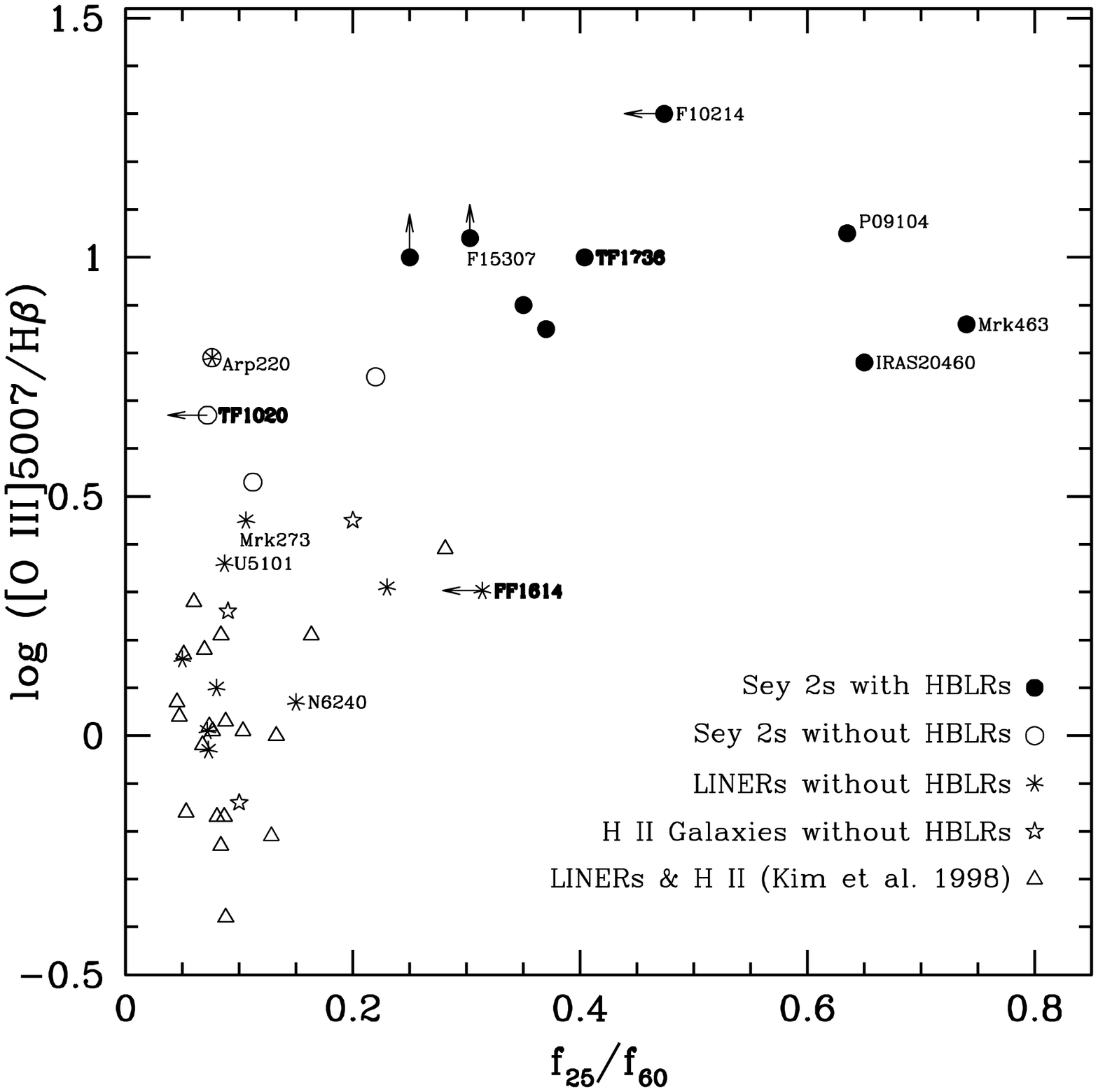} 
\caption{{\it Left panel:} Spectropolarimetry of TF~J1736+1122. (a) Total flux spectrum, (b) observed degree of polarization, and (c) polarized flux spectrum.
{\it Right panel:} \oiii~\wave 5007/\hb~versus IR color \firr~for narrow-line 
ULIRGs in which HBLRs have been searched for. Also plotted in open triangles 
are all ULIRGs classified as H II and LINERs in Kim et al. (1998). Note the 
clear tendency for ULIRGs with HBLRs to have warmer IR color and higher 
excitation spectrum. The opposite holds for LINERs and H II galaxies.} 
\end{figure} 

It is of interest to see if it could be 
determined, from optical spectroscopic and $IRAS$ photometric data alone,
whether an ULIRG harbors a genuine quasar or is powered by a starburst.
Figure 1b shows a plot of the line ratio \oiii~\wave 5007/\hb(narrow),
which can serve as an indicator of the ionization level, versus the infrared
color \firr. As can be seen, there is a clear tendency for higher-ionization 
and warmer Seyfert 2 ULIRGs to show hidden broad-line region (HBLR) indicative of a ``buried quasar.''
Furthermore, Seyfert ULIRGs without HBLRs lie in a region of the diagram
similar to that occupied by H II and LINER galaxies, none of which has been
found to have HBLRs. 

\acknowledgments 
Research at IGPP/LLNL is performed under the auspices of the 
US Department of Energy under contract W-7405--ENG--48.


\scriptsize
\begin{references}
\reference Becker, R. H., White, R. L., \& Helfand, D. J. 1995, \apj, 450, 559
\reference Condon, J. J. et al. 1991, \apj, 378, 65 
\reference Dey, A., \& van Breugel, W. 1994, in Mass-Transfer Induced Activity in Galaxies, ed. I. Shlosman, p. 263
\reference Douglas, J. N. et al. 1996, \aj, 111, 1945
\reference Kim, D.-C., Veilleux, S., \& Sanders, D. B. 1998, \apj, in press, astro-ph/9806149
\reference Sanders, D. B. et al. 1988, \apj, 325, 74
\reference Soifer et al. 1987, \apj, 320, 238
\reference Stanford, S. A., Stern, D., van Breugel, W., De Breuck, C. 1998, \apj, in prep.
\end{references}
\end{document}